\documentclass{PoS}

\usepackage{amsfonts}
\usepackage{amsmath}
\usepackage{subfigure} 

\graphicspath{{./figs/}}

\title{
  Systematic Uncertainties in $B_K$ with Improved Staggered Fermions
}

\ShortTitle{
  Systematic Uncertainties in $B_K$
}

\author{\speaker{Yong-Chull Jang}, Taegil Bae,
  Hyung-Jin Kim, Jangho Kim,
  Jongjeong Kim, Kwangwoo Kim, Boram Yoon, Weonjong Lee\\
  Lattice Gauge Theory Research Center, CTP, and FPRD, \\
  Department of Physics and Astronomy,
  Seoul National University, Seoul, 151-747, South Korea \\
  E-mail: \email{wlee@snu.ac.kr}}

\author{Chulwoo Jung \\
  Physics Department, Brookhaven National Laboratory,
  Upton, NY11973, USA \\
  E-mail: \email{chulwoo@bnl.gov}}

\author{Stephen R. Sharpe\\
  Physics Department, University of Washington, Seattle, WA 98195-1560 \\
  E-mail: \email{sharpe@phys.washington.edu}}

\abstract{ We study three sources of error in 
our calculation of $B_K$ using HYP-smeared staggered fermions on
the MILC asqtad lattices. These are (1) dependence
  on the light sea quark mass; (2) finite volume effects; and
(3) the impact of an order of magnitude increase in the number
of measurements. Our main results are 
(1) the dependence on the light sea-quark mass
is weaker than expected by naive dimensional analysis,
(2) including finite volume effects in SU(2) staggered chiral
perturbation theory fits leads to a very small change in $B_K$,
of size $\approx 0.1\%$, and (3) increasing the statistics
on one of the coarse MILC lattices resolves a potential
discrepancy with other coarse results.
}

\FullConference{
  The XXVIII International Symposium on Lattice Field Theory, Lattice2010\\
  June 14-19, 2010\\
  Villasimius, Italy
}

\begin{document}

\section{Introduction} 
This paper is the third in a series of four proceedings 
describing our calculation of $B_K$ using improved staggered fermions.
Here, we review some of the errors quoted in the
error budgets for $B_K$ given in the companion
proceedings \cite{ref:wlee-2010-2} and \cite{ref:wlee-2010-3}.
In particular, we consider the following issues:
\begin{itemize}
\item The dependence of $B_K$ on the light sea quark mass $am_\ell$;
\item Finite volume effects in the SU(2) analysis of $B_K$;
\item The effect of increasing the number of measurements on the C4 ensemble.
\end{itemize}
For our notations for fits, and details of the ensembles we use,
see Refs.~\cite{ref:wlee-2010-2,ref:wlee-2010-3}, as well
as our recent long article~\cite{ref:wlee-2010-1}.

\section{Dependence of $B_K$ on light sea-quark masses}
A year ago, we studied the light sea-quark mass dependence
using five different ``coarse'' ($a\approx 0.12\;$fm) MILC ensembles, with 
results presented (using fits based on 
SU(2) staggered chiral perturbation theory [SChPT])
in Ref.~\cite{ref:wlee-2009-2}.
In the intervening year, we have added a second sea-quark mass on the
fine ($a\approx 0.09\;$fm) lattices (ensemble F2, with $am_\ell:a m_s=0.0031:0.031$,
i.e. with $m_\ell$ halved compared to ensemble F1),
and also increased the statistics on one of the coarse ensembles
(C4, with $a m_\ell: a m_s=0.007:0.05$).
This allows us to solidify our understanding of the sea-quark mass
dependence. This section summarizes the more extensive discussion
given in Ref.~\cite{ref:wlee-2010-1}.

Both SU(2) and SU(3) analyses allow us to extrapolate
our results to the physical valence $d$ and $s$ masses,
assuming that the corresponding ChPT is convergent.
We use the SChPT fit forms to correct the chiral logarithms
for the fact that the sea-quark masses differ from their
physical values. Once these logarithmic corrections
have been accounted for, 
the remaining dependence on sea quark masses is analytic,
and given by
\begin{equation}
B_K = b_0 + b_1 (am_\ell) + b_2 (a m_s) + {\cal O}\left[(am)^2\right]\,.
\label{eq:BKvsam}
\end{equation}
Here $a m_\ell$ and $a m_s$ are the light and strange bare sea-quark masses,
respectively.
In SU(3) ChPT we have the additional relation $b_2=b_1/2$, while
in SU(2) ChPT $b_2$ and $b_1$ are unrelated.
In practice, we can only determine $b_1$, since, to date, all our
ensembles have, for a given lattice spacing, the same value of $a m_s$.

\begin{figure}[tbhp]
\centering
\includegraphics[width=0.47\textwidth]{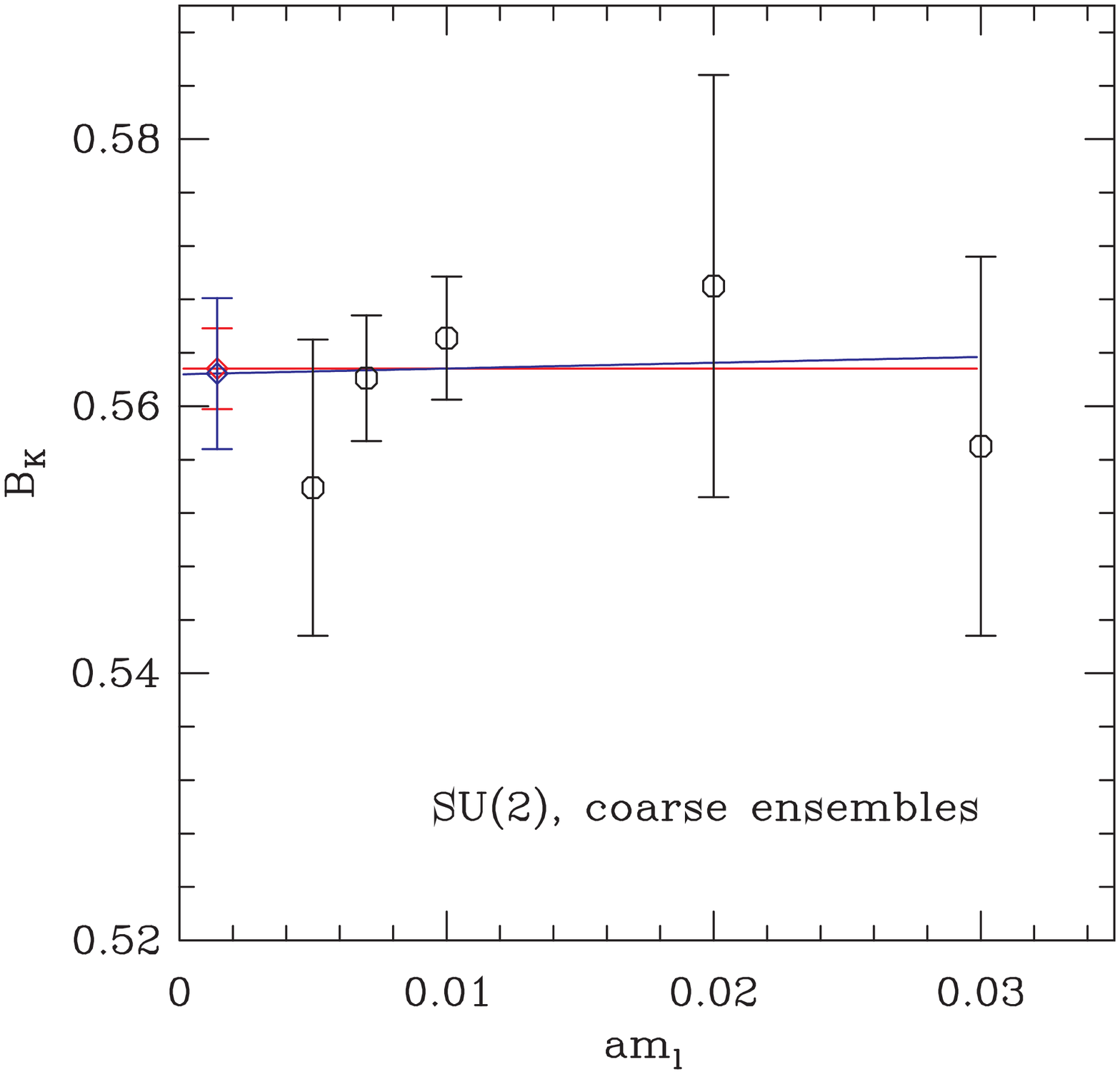}
\includegraphics[width=0.49\textwidth]{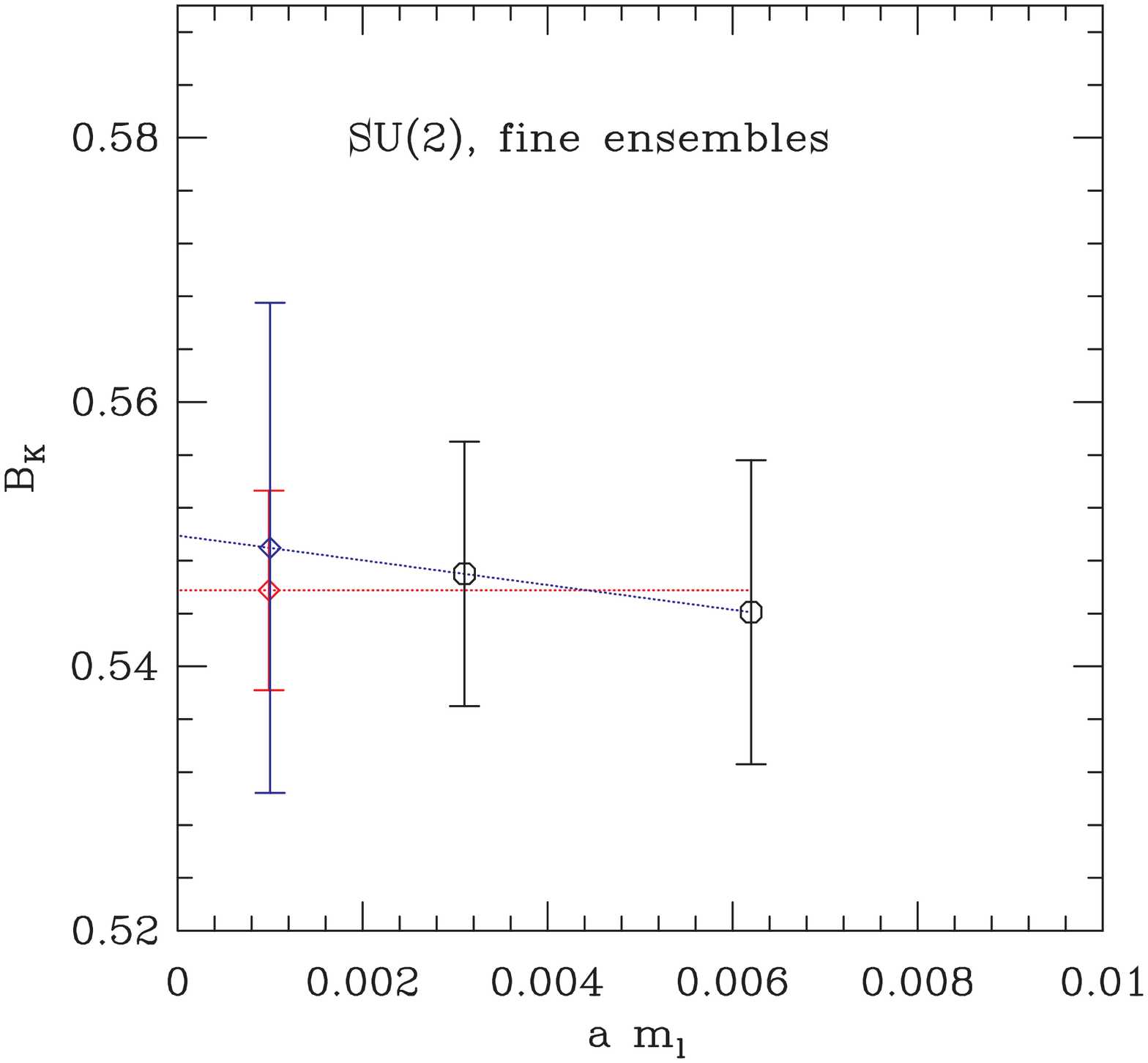}
\caption{$B_K$ vs. $am_\ell$ for the MILC coarse ensembles (left) and
  for the MILC fine ensembles (right). These values are obtained using
 ``4X3Y-NNLO'' SU(2) fits~\protect\cite{ref:wlee-2010-1,ref:wlee-2010-2}.
 $B_K$ is obtained using one-loop matching and
 is evaluated at the scale $\mu=2$ GeV.
}
\label{fig:su2-4x3y-nnlo-ml}
\end{figure}
In Fig.~\ref{fig:su2-4x3y-nnlo-ml}, we show the dependence
of $B_K$ (after extrapolation to physical valence quark masses)
on $a m_\ell$, for both coarse and fine lattices.
The red and blue lines show, respectively, fits to a constant and a linear
function. The fit parameters are given in Table~\ref{tab:su2:am_l}.
We see that the slope $b_1$ is consistent with zero,
but has rather large errors. 
To compare the slopes at the two lattice spacings, we rewrite the
expression~(\ref{eq:BKvsam}) as
\begin{equation}
B_K = b_0 \left[1 +   b'_1 (L_P/\Lambda_\chi^2) \right]\,,
\end{equation}
where $L_P$ is the mass-squared of the pion composed of light
valence quarks (in physical units),
and $\Lambda_\chi$ is the expansion scale of ChPT, which
we take to be $1\;$GeV.
Expressed this way, the slope coefficient
$b'_1$ should be the same for both lattice spacings, up to
(presumably small) discretization errors.
Furthermore, naive dimensional analysis suggests that 
$|b'_1| ={\cal O}(1)$.
Values for $b'_1$ are also given in the Table,
and show that the magnitudes of the slopes are, in fact,
considerably smaller than expected. This is not problematic,
since a small value will occur some of the time.
It is, however, serendipitous, since
it reduces the uncertainty in the extrapolation to physical
$a m_\ell$.

%
%
%
%
\begin{table}[h!]
\begin{center}
\begin{tabular}{c | c | c | c}
\hline
\hline
$a$ (fm) & $b_0$ & $b_1$ & $b'_1$ \\ 
\hline
0.12 &  0.5624(64) & $+$0.04(58) & $+$0.005(74)\\
0.09 &  0.550(23)  & $-$0.93(492) & $-$0.09(46) \\
\hline
\hline
\end{tabular}
\end{center}
\caption{Parameters of the linear fits to SU(2) results shown in 
Fig.~\protect\ref{fig:su2-4x3y-nnlo-ml}.
}
\label{tab:su2:am_l}
\end{table}
\begin{figure}[tbhp]
\centering
\includegraphics[width=0.49\textwidth]{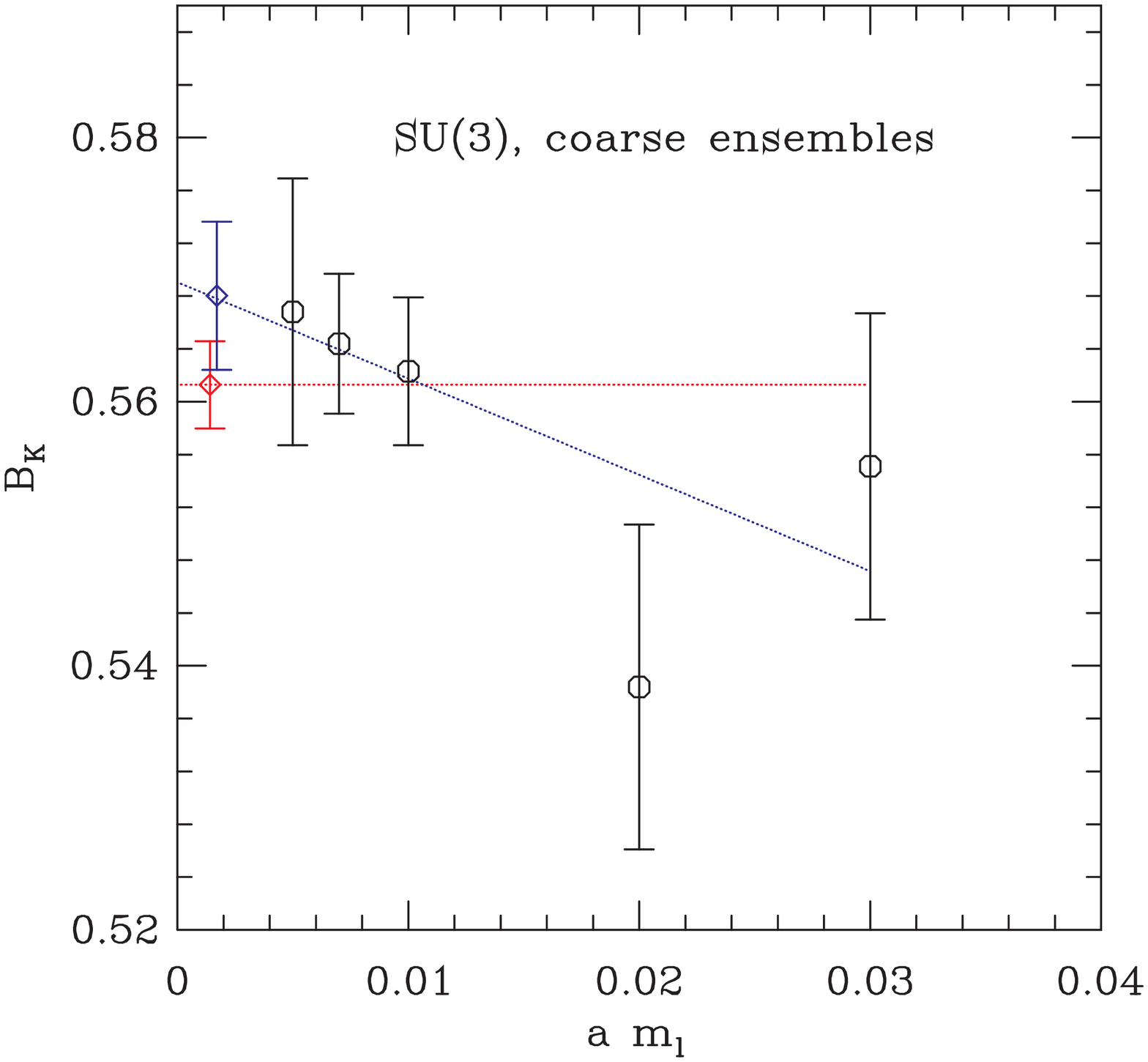}
\includegraphics[width=0.49\textwidth]{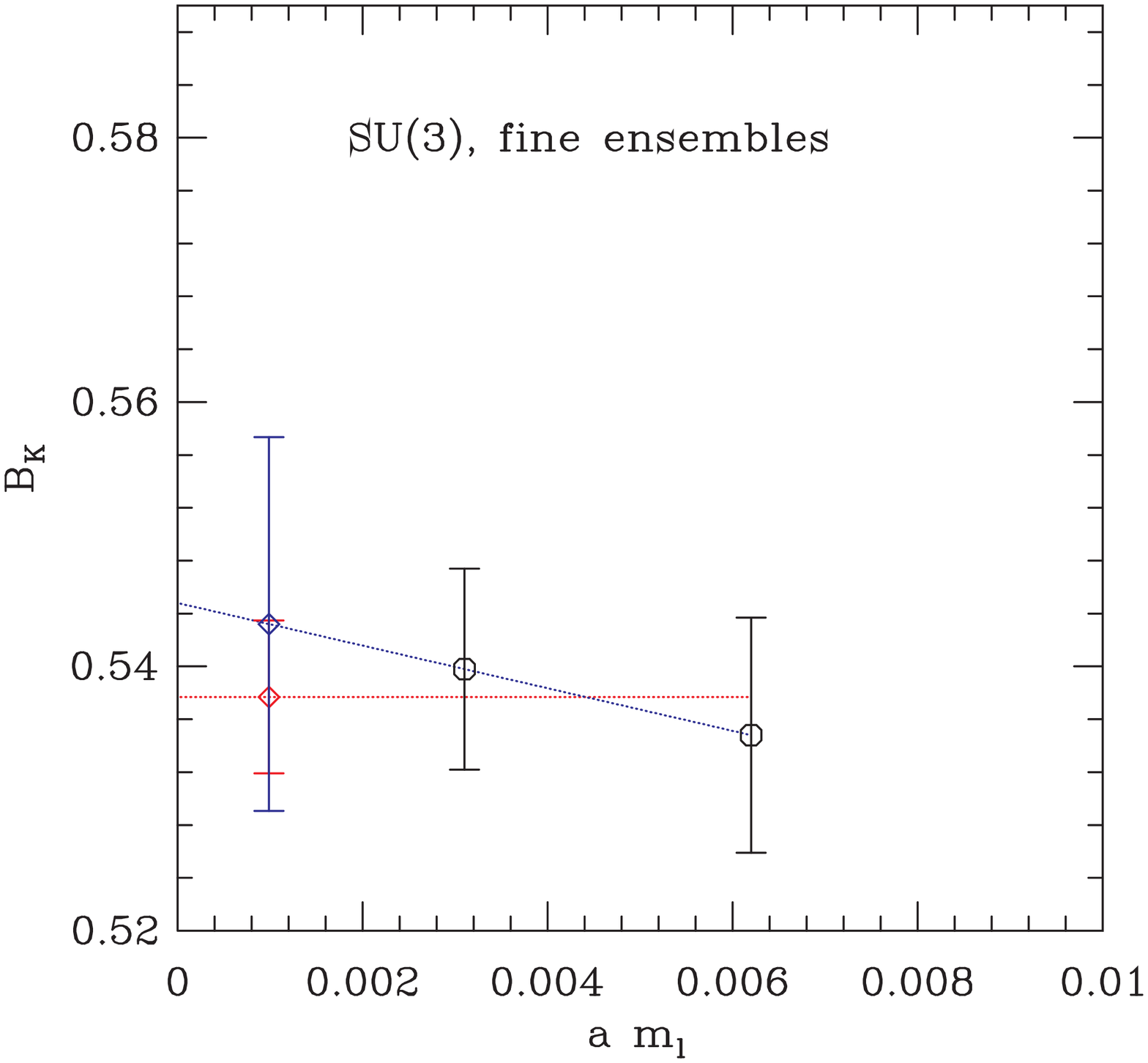}
\caption{As in Fig.~\protect\ref{fig:su2-4x3y-nnlo-ml},
but for the SU(3) analysis, using N-BB1 fits 
(see Ref.~\protect\cite{ref:wlee-2010-3}).
}
\label{fig:su3-nbb1-ml}
\end{figure}
The corresponding fits for the SU(3) SChPT analysis
are shown in Fig.~\ref{fig:su3-nbb1-ml} 
with parameters given in Table~\ref{tab:su3:am_l}.
They give central values having somewhat larger magnitudes than the
SU(2) fits, 
but they are still smaller than expected from naive dimensional analysis.
Note that the slopes from the SU(2) and SU(3) fits need not be the
same.
%
%
\begin{table}[h!]
\begin{center}
\begin{tabular}{c | c | c | c}
\hline
\hline
$a$ (fm) & $b_0$ & $b_1$ & $b'_1$ \\ 
\hline
0.12 &  0.5691(62)  & $-$0.73(49)   & $-$0.09(6)\\
0.09 &  0.5448(176) & $-$1.61(3.78) & $-$0.15(36)\\
\hline
\hline
\end{tabular}
\end{center}
\caption{Parameters of the linear fits to SU(3) results 
shown in Fig.~\protect\ref{fig:su3-nbb1-ml}.
}
\label{tab:su3:am_l}
\end{table}

Since we find no evidence for a significant dependence on
$m_\ell$, we assume, for our final value of $B_K$, that there
is no such dependence, and do the continuum extrapolation
using the $m_\ell/m_s=1/5$ lattices (including C3 and F1). 
We then correct for a possible dependence on $a m_\ell$ by 
including in the error budgets a systematic error which
is the difference in $B_K$ between the result on the C3 ensemble
and that obtained using linear extrapolation to $a m_\ell^{\rm phys}$
\cite{ref:wlee-2010-1,ref:wlee-2010-2,ref:wlee-2010-3}.
The error from an incorrect value of $a m_s^{\rm phys}$ is 
estimated similarly.

\section{Finite Volume Effects from SU(2) SChPT}
Finite volume (FV) dependence is predicted by ChPT. At NLO, this dependence
enters through corrections to the chiral logarithmic functions, as follows:
\begin{eqnarray}
\ell(X) &=& X \left[\log(X/\mu_{\rm DR}^2) +\delta^{\rm FV}_1(X) \right]\,, 
\label{eq:app:l}
\\
\tilde\ell(X) &=& - \frac{d\ell(X)}{dX}
=
-\log(X/\mu_{\rm DR}^2) -1 +\delta^{\rm FV}_3(X)\,.
\label{eq:app:tilde-l}  
\end{eqnarray}
Here $X$ is the mass-squared of a pion,
and $\mu_\textrm{DR}$ is the scale introduced by dimensional
regularization.\footnote{%
The NLO result is independent of $\mu_\textrm{DR}$ once
one includes the analytic terms, and in any case 
$\mu_\textrm{DR}$ does not enter the FV corrections,
$\delta^{\rm FV}_{1,3}$.
}
The FV corrections have the form of an image sum,
\begin{equation}
\delta^{\rm FV}_1(M^2) = \frac4{ML} \sum_{\vec n\ne 0}
\frac{K_1(|\vec n| ML)}{|\vec n|}\,,
\qquad
\delta^{\rm FV}_3(M^2) = 2 \sum_{\vec n\ne 0}
{K_0(|\vec n| ML)}\,,                                                        
\end{equation}
where $L$ is the box size, and $\vec n$ labels the image position.
For our geometries, sufficient accuracy
is attained by keeping only spatial images.

In our previous work, including our long article~\cite{ref:wlee-2010-1},
we have not included these finite volume corrections, 
due to the high computational cost of implementing them.
Instead, we have shown that the expected size of these corrections is small
compared to other errors~\cite{ref:wlee-2009-3}.
This year, however, we have been able to do one-loop SU(2) 
SChPT fits including FV effects
(keeping sufficient images that no approximation is made at double
precision accuracy).
To do so has required that we use GPUs.
On a single core of the Intel i7 920 CPU 
(running at $\sim 0.5$ giga flops per core), 
a single FV fit to all ten MILC ensembles that we use takes about a week.
Using an Nvidia GTX 480 GPU, by contrast,
we have obtained a sustained performance of 67 giga flops
(45\% of the peak speed)~\cite{ref:wlee-2010-6}.
Hence, using the GTX 480 GPU, it takes only about an hour to do the full
SU(2) analysis for all the MILC ensembles.
This is fast enough to carry out multiple fits, as is needed to
estimate fitting errors.

\begin{figure}[t!]
\centering
\includegraphics[width=0.49\textwidth]
{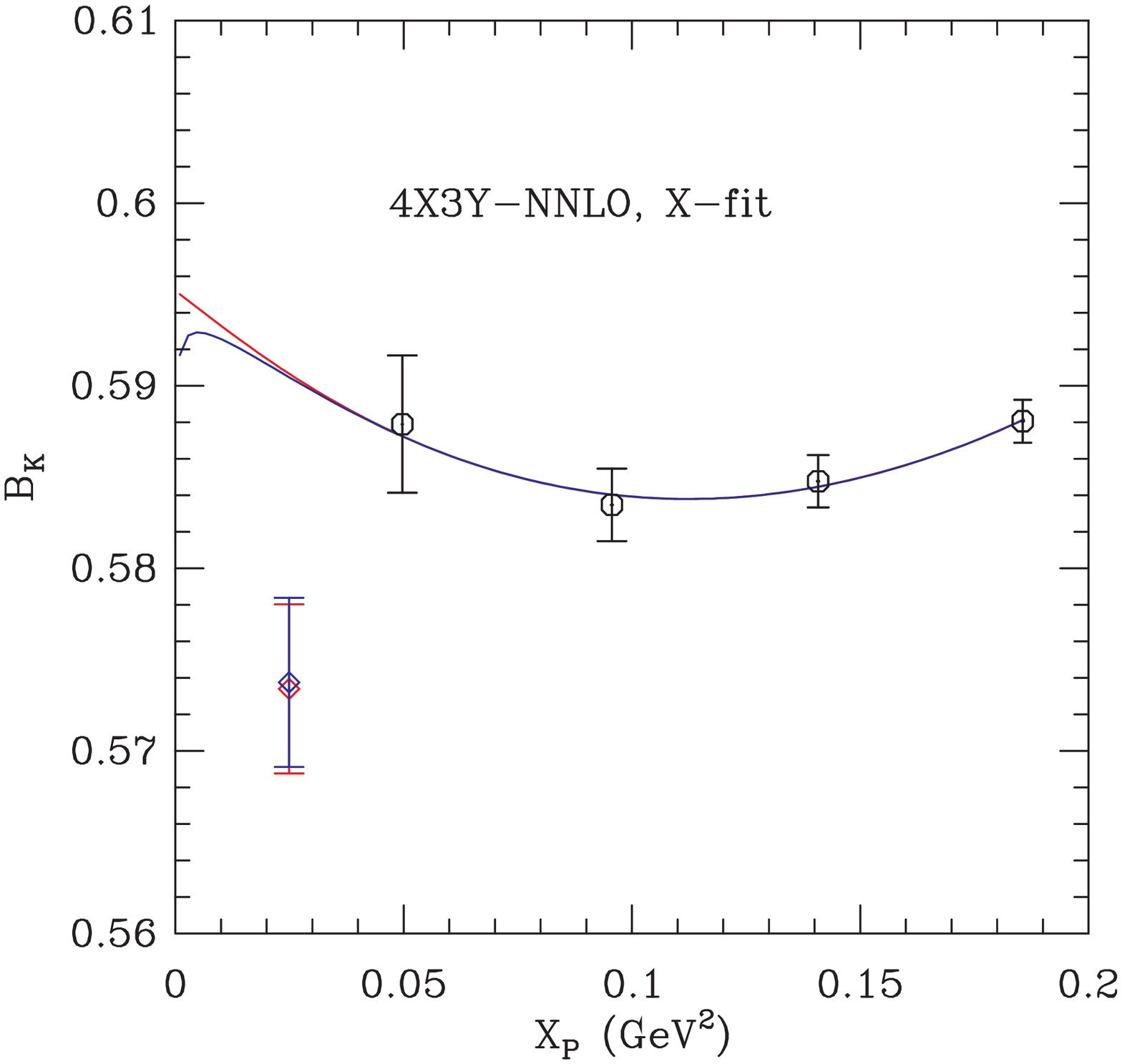}
\includegraphics[width=0.49\textwidth]
{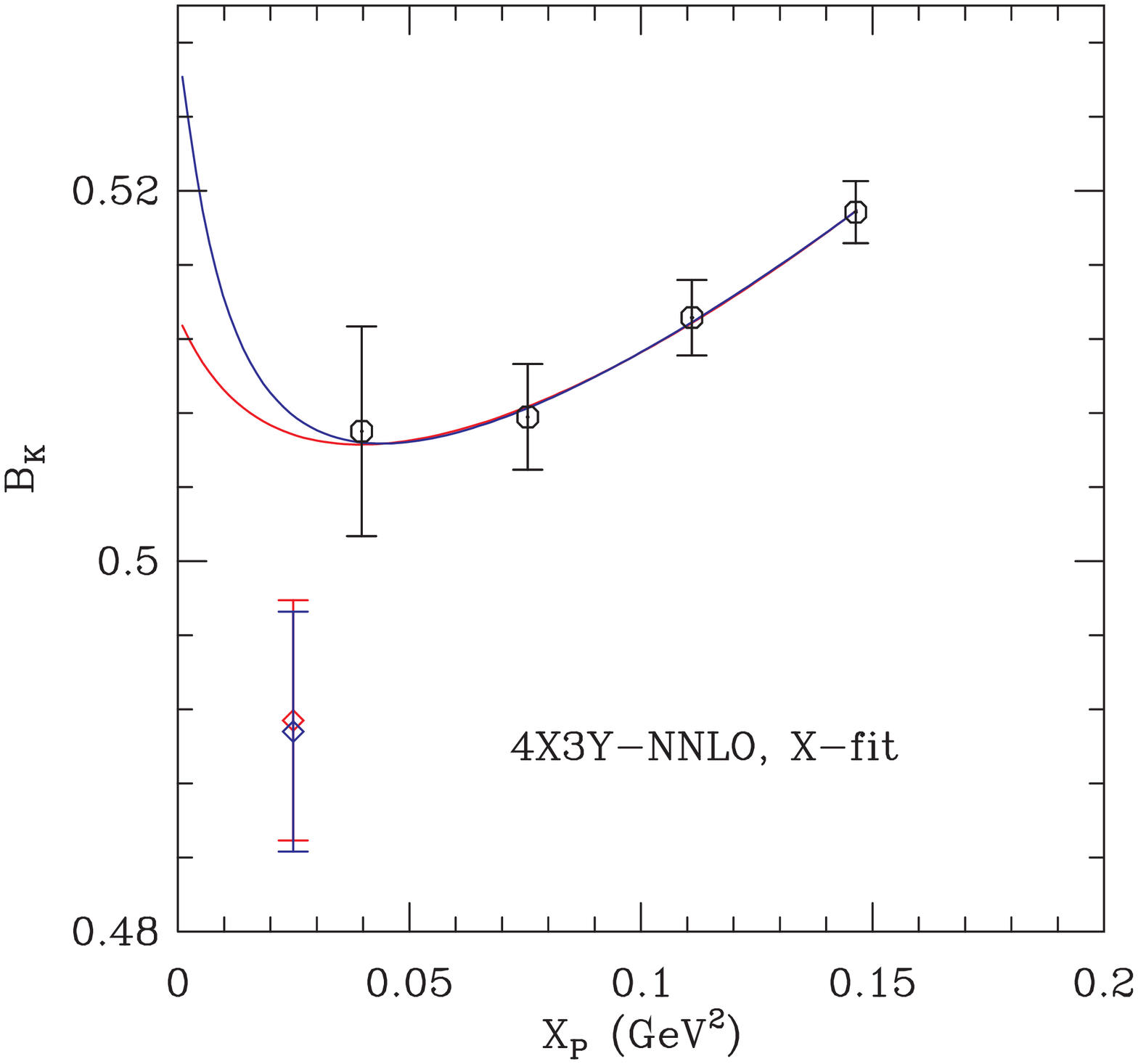}
\caption{$B_K(\mu=1/a)$ (one-loop matched) versus $X_P$ (mass-squared
  of Goldstone-taste pion composed of quark and antiquark of mass
  $m_x$), including 4X-NNLO fits (explained in
  Ref.~\protect\cite{ref:wlee-2010-1}).  Ensembles are C3 (left, with $a
  m_y=0.05$) and S1 (right, with $a m_y=0.018$) ensembles.  The red (blue)
  lines shows fit functions without (with) FV corrections included.  See
  text for more details.
}
\label{fig:su2-4x3y-nnlo:FV}
\end{figure}
In Fig.~\ref{fig:su2-4x3y-nnlo:FV}, we compare the fits with and without
FV corrections on a coarse (C3) and a superfine (S1--$a\approx 0.06\;$fm) 
ensemble.
The red line shows the $V\rightarrow\infty$ fit, while the
blue line includes FV corrections.
The results of fits for $B_K$ are quoted in Table~\ref{tab:su2-fv}. 
These correspond to the red and blue points in the figures,
and have been obtained by setting $m_d$ and $m_\ell$ to their
physical values in the fit function, as well as removing taste
splittings and setting $V\to\infty$ in the FV form.
The impact of using the FV fit form is very small ($\sim 0.1\%$),
as can be expected from the fact that
the fit functions hardly differ for our pion masses.
It is only for smaller values of $X_P$ that FV effects are visible.
Note that the FV corrections
have opposite signs on the two ensembles.
This is due to a competition between two terms having
different dependence on taste-splittings (and thus having different
size on the two ensembles)~\cite{ref:wlee-2009-3}.

\begin{table}[h!]
\begin{center}
\begin{tabular}{c | c | c }
\hline
\hline
ID & $B_K$ & $B_K$(FV) \\
\hline
C3 & 0.5734(46) & 0.5738(46) \\
S1 & 0.4914(65) & 0.4908(65) \\
\hline
\hline
\end{tabular}
\end{center}
\caption{Results for $B_K(\mu=1/a)$ with and without
the finite volume corrections in the fit functions. 
The fits are those shown in Fig.~\protect\ref{fig:su2-4x3y-nnlo:FV}.
}
\label{tab:su2-fv}
\end{table}
It is well known that the one-loop prediction of FV effects
gives only a semi-quantitative guide to their magnitude, since
higher-loop effects can be important.
For this reason, and to be conservative, we do not use the results
just presented to estimate the FV systematic.
Instead, we use the difference in $B_K$ obtained on
the C3 and C3-2 ensembles (which have different spatial volumes),
which leads to an error estimate of 0.85\%~\cite{ref:wlee-2010-1}.

\section{Effect of Higher Statistics}
Since last year, we added an additional 9 measurements on each lattice
of the C4 ensemble, with the source timeslices chosen randomly, and
different random seeds for the wall sources.
In Ref.~\cite{ref:wlee:2009-4} we found that such additional measurements
are, to good approximation, statistically independent.
This new data
 allows us to resolve a small puzzle we had observed in last year's data.

\begin{figure}[t!]
\centering
\includegraphics[width=0.49\textwidth]
{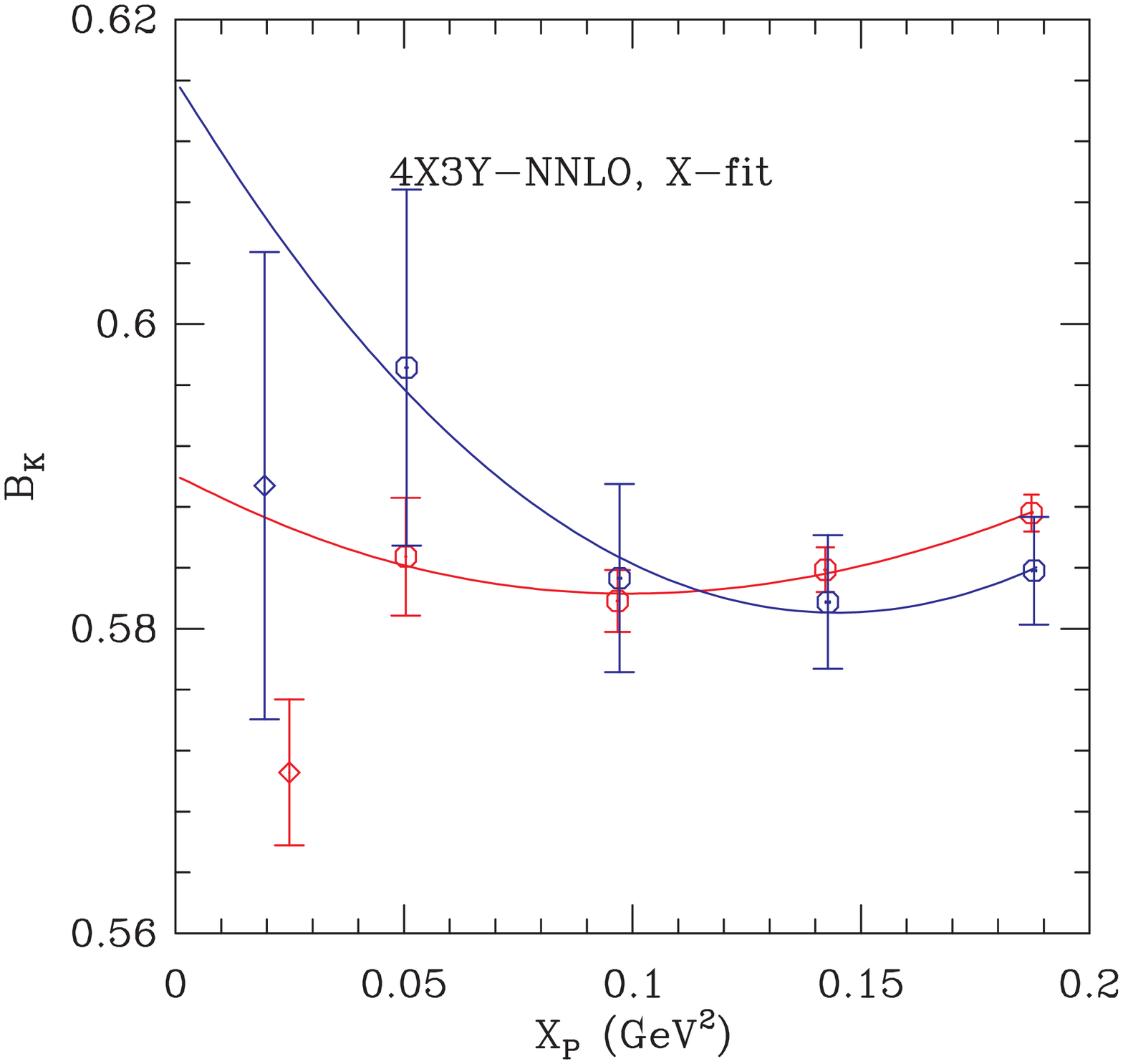}
\includegraphics[width=0.49\textwidth]
{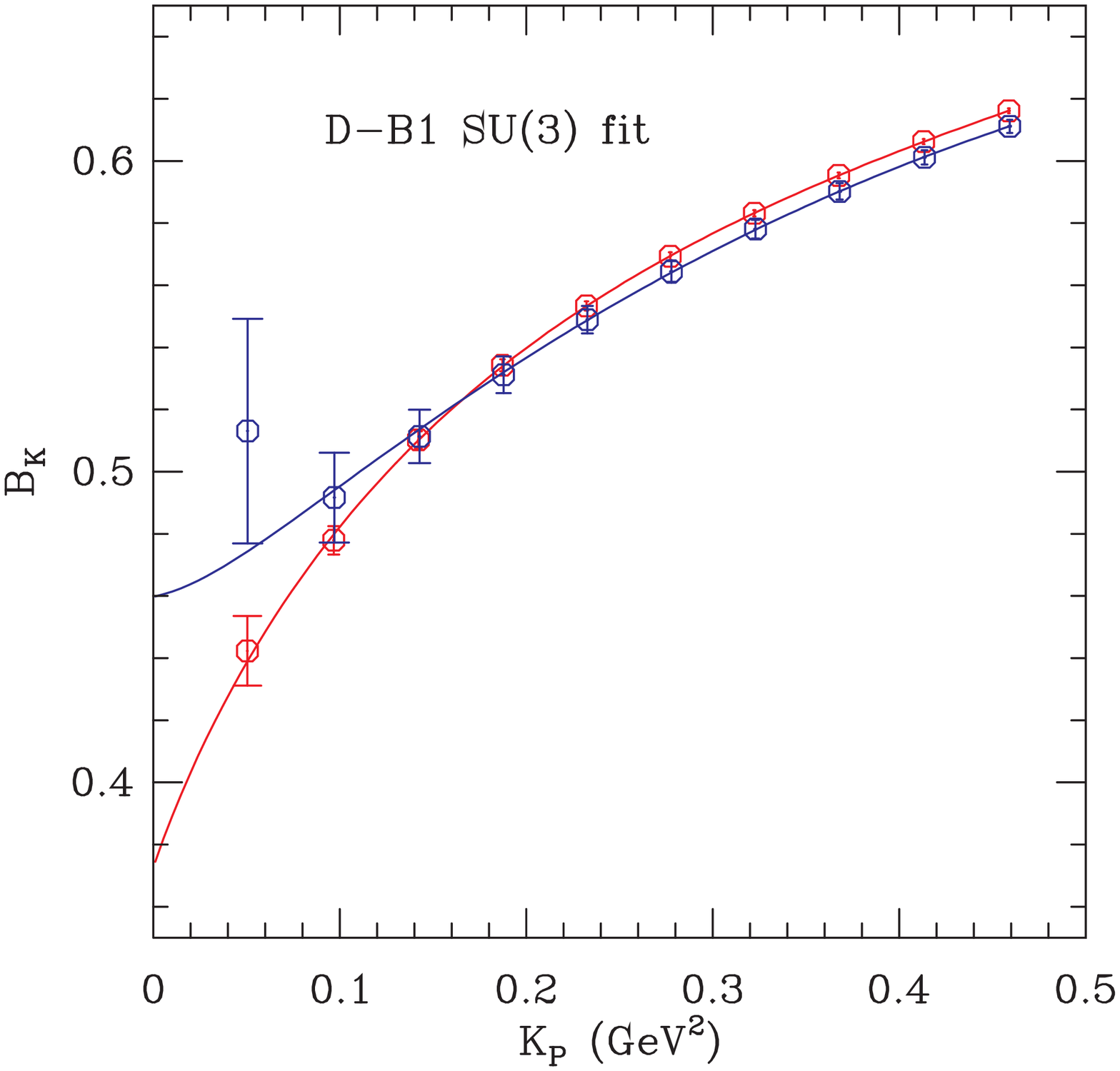}
\caption{ $B_K(\text{NDR}, \mu=1/a)$ as a function of $X_P$
  for the SU(2) analysis (left) and the SU(3) analysis (right) on
  the C4 ensemble. 
  The fit types are 4X3Y-NNLO for SU(2) and D-B1 for SU(3)
  (see Ref.~\protect\cite{ref:wlee-2010-1} for details). 
  Blue points and fit correspond to 1 measurement/config
  while red data points and fit correspond to 10 measurements/config.
}
\label{fig:bk:stat:C4}
\end{figure}

In Fig.~\ref{fig:bk:stat:C4}, we compare the low and high statistics data.
The data points have shifted in a way which is 
consistent with their original statistical uncertainties, 
but the result is a behavior, particularly at low $X_P$,
more similar to that observed on other ensembles.
Examples of the behavior on other ensembles are shown in 
Refs.~\cite{ref:wlee-2010-1,ref:wlee-2010-2,ref:wlee-2010-3}.

This result illustrates the importance of obtaining small
statistical errors, and also shows again how
using multiple measurements on a single configuration
is an efficient way of reducing errors.

\section{Acknowledgments}
C.~Jung is supported by the US DOE under contract DE-AC02-98CH10886.
The research of W.~Lee is supported by the Creative Research
Initiatives Program (3348-20090015) of the NRF grant funded by the
Korean government (MEST). 
The work of S.~Sharpe is supported in part by the US DOE grant
no.~DE-FG02-96ER40956.
Computations were carried out in part on QCDOC computing facilities of
the USQCD Collaboration at Brookhaven National Lab. The USQCD
Collaboration are funded by the Office of Science of the
U.S. Department of Energy.

\end{document}